\documentclass[preprint, amssymb,  aps, prl]{revtex4-1}

\usepackage{graphicx, subfigure}
\usepackage{bm}
\usepackage{times}
\usepackage{amsfonts}
\usepackage{amsmath}
\usepackage{bm}
\usepackage{amssymb}
\usepackage{amsfonts}
\usepackage{amsmath}

\begin{document}

%\begin{frontmatter}

\title{The role of the angular momentum of light in Mie scattering. Excitation of dielectric spheres with Laguerre-Gaussian modes}

%% use optional labels to link authors explicitly to addresses:
%% \author[label1,label2]{<author name>}
%% \address[label1]{<address>}
%% \address[label2]{<address>}

\author{Xavier Zambrana-Puyalto and Gabriel Molina-Terriza}

\address{QsciTech and Department of Physics and Astronomy, Macquarie University, Australia}
\address{ARC Center of Excellence for Engineered Quantum Systems (EQuS)}

\begin{abstract}
We present a method to enhance the ripple structure of the scattered electromagnetic field in the visible range through the use of Laguerre-Gaussian beams. The position of these enhanced ripples as well as their linewidths can be controlled using different optical beams and sizes of the spheres. 
\end{abstract}

%\begin{keyword}
%Laguerre-Gaussian \sep Mie \sep resonances \sep scattering \sep multipole 
%\end{keyword}

\maketitle 
%\end{frontmatter}

\section{Introduction}

In 1890, Ludvig Lorenz obtained one of the few analytical solutions of the Maxwell equations in inhomogeneous media, the scattering of a plane wave from a dielectric sphere \cite{Lorenz1898}. Later, Gustav Mie rediscovered this result and applied it to metallic spheres \cite{Mie1908}. His results explained the change in colors of colloidal suspensions of gold nanoparticles of different sizes, also showing for the first time the size dependence of localized plasmon resonances in nanostructures. The resulting theory is usually called Lorenz-Mie theory (or Mie Theory) and solves the scattering problem of an incident plane wave propagating in a homogeneous isotropic medium hitting a homogeneous isotropic sphere. However, with the advent of modern computation techniques, this theory has been developed further to the Generalized Lorenz-Mie theory (GLMT), which solves the scattering problem for any incident electromagnetic field \cite{GLMT_book}. The GLMT has found application in many and diverse fields. We refer the reader to Gouesbet \textit{et al.} \cite{Gouesbet2011} for an extensive review of all the possible applications. In particular, it is important to highlight the remarkable impact that this theory has had in the development of nanophotonics. Some very recent examples are the study of photonic jets \cite{Devilez2008}, the characterization of coherent perfect absorbers \cite{Noh2012}, the prediction of the optical pulling force \cite{Chen2011} or the control of localized surface plasmons \cite{Mojarad2008}, or the excitation of multipolar resonances \cite{Zambrana2012}.

%from measurement techniques such as Phase-Doppler instruments and rainbow refractometry, to optical manipulation of particles as well as control of internal fields in certain structures. 

However, one aspect of the GLMT that is usually overlooked is its intimate relation to the angular momentum (AM) of electromagnetic waves \cite{Jackson1998}. The GLMT is naturally described in spherical coordinates, since the scatterer is a sphere.  The multipolar modes $ \mathbf{A}_{jm}^{(x)} $ are the solutions of Maxwell equations in spherical coordinates where $j=1,2,..,\infty$ stands for the angular momentum of the mode, $\vert m \vert \le j$ is the third component of the angular momentum and $x=e,m$ accounts for the parity of the beam, $m$ being the magnetic multipole and $e$ the electric one. Indeed, it can be proved that $J^2 \mathbf{A}_{jm}^{(x)} = j(j+1) \mathbf{A}_{jm}^{(x)} $, $J_z \mathbf{A}_{jm}^{(x)} = m \mathbf{A}_{jm}^{(x)}$, $P \mathbf{A}_{jm}^{(m)} = (-)^j \mathbf{A}_{jm}^{(m)}$ and $P \mathbf{A}_{jm}^{(e)} = (-)^{j+1} \mathbf{A}_{jm}^{(e)}$. That is, the multipolar modes are eigenstates of the AM squared operator $J^2$, the $z$ component of the AM operator $J_z$, and the parity operator $P$. The field of AM of light was significantly developed when in 1992 Allen and co-workers showed that in the paraxial approximation one could find a set of modes which had a well defined value of the third component of the AM \cite{Allen1992}, $J_z$. A particular set of paraxial modes with this property are the Laguerre-Gaussian modes (LG$_{q,l}$, with $q$ the radial index and $l$ the azimuthal one). This important finding opened up a whole new field allowing for innumerable applications in quantum optics, microscopy, biology, optical trapping and astrophysics, just to mention a few \cite{Sonja2008}. Finally, although the total AM of a vectorial field is composed of an orbital part ($\mathbf{L}$) and a spin part ($\mathbf{S}$), i.e. $\mathbf{J}=\mathbf{L}+\mathbf{S}$, these two components cannot physically be separated. That is, they can be split mathematically, but they do not correspond to any physical observable. When either of these operators are applied to a Maxwell field, the result from that operation is not a Maxwell field \cite[p50]{Cohen1997}\cite[\S 16]{Lifshitz1982}.

In this article we explore the use of LG modes to excite spherical dielectric nanoparticles. Exploiting the properties of angular momentum and using some analytical techniques we manage to enhance the ripple structure of an arbitrary sphere in a lossless, homogeneous and isotropic medium. %A similar computational study using metallic particles was done in \cite{vandeNes07}. Nevertheless, the resonances described in this paper were hidden therein due to the absorption of the metallic particles. 
Furthermore, the results presented in this paper will allow for a better understanding of recent scattering experiments of silica spheres with Laguerre-Gaussian beams \cite{Gabi2012OL}.

\section{Generalized Lorenz-Mie Theory}

In this section we solve the following scattering problem: a monochromatic LG$_{q,l}$ beam propagating in a homogeneous, isotropic and lossless me\-dium impinges on a dielectric sphere made of a homogeneous and isotropic material. Since all the material properties can be reduced to a pair of scalar numbers (the electrical permittivity $\epsilon$ and the magnetic permeability $\mu$) for both the sphere and the surrounding medium, the electomagnetic field can be obtained in two steps. First, one of the two fields ($\mathbf{E}$ or $\mathbf{H}$) is found by solving the monochromatic wave equation. Then, the other field is obtained by applying the curl to the first obtained field:
\begin{eqnarray}
\nabla^2 \mathbf{E} + k^2 \mathbf{E} &=& 0\\
\nabla \times \mathbf{E} &=&ik\mu \mathbf{H}
\end{eqnarray}
where $k=n \omega/c$ with $n=\sqrt{\epsilon \mu}$ the index of refraction of the medium. The temporal dependence of the fields is supposed to be $\exp(-iwt)$ in the whole text.  The multipolar fields $\left\lbrace \mathbf{A}_{jm}^{(m)},  \mathbf{A}_{jm}^{(e)} \right\rbrace$  are precisely a basis of solution for the two equations above \cite{Rose1955}. Furthermore, as mentioned in the introduction, they are specially well suited for problems in spherical coordinates as they are rotationally symmetric. Thus, we will use them as a basis to decompose all the fields in the problem. The fields we have to consider are the incident ($\mathbf{E}^i$, $\mathbf{H}^i$), the scattering ($\mathbf{E}^{sca}$, $\mathbf{H}^{sca}$) and the interior field ($\mathbf{E}^{l}$, $\mathbf{H}^{l}$). The incident field is given (it is a LG$_{pl}$ in this case), therefore its decomposition in multipoles can be computed. In general, this decomposition cannot be evaluated analytically. That is, if the incident field has a decomposition of the form $\mathbf{E}^{i}=\sum_{j,m}g_{jm}^{(m)}\mathbf{A}_{jm}^{(m)} + g_{jm}^{(e)}\mathbf{A}_{jm}^{(e)}$ then both $g_{jm}^{(m)}$ and $g_{jm}^{(e)}$ are obtained by computing two cumbersome double integrals $g_{jm}^{(x)}= \int_0^{\pi} \int_0^{2\pi}  \mathbf{E}^i \cdot \mathbf{A^*}_{jm}^{(x)} \sin \theta d\theta d\phi $. However, once this has been done, it has been proved that the scattered and the interior fields are readily determined by applying Maxwell boundary conditions on the sphere \cite{GLMT_book}. Thus, the only technicality of the problem is finding the decomposition of the incident LG$_{q,l}$ beam into multipoles. This decomposition was found in \cite{Gabi2008}:
\renewcommand{\arraystretch}{2.5}
\begin{equation}
\begin{array}{ccl}
\textbf{A}&= &\displaystyle\sum_{j=\vert l+p \vert}^{\infty} i^j (2j+1)^{1/2} C_{jlp} \left[ \textbf{A}_{j(l+p)}^{(m)}+ip\textbf{A}_{j(l+p)}^{(e)} \right] \\
C_{jlp}&= & \displaystyle\int_0^{\pi}\sin\theta d_{(l+p)p}^j(\theta)f_{q,l}(k\sin\theta) \, d\theta 
\end{array}
\label{general}
\end{equation}
\renewcommand{\arraystretch}{1} 
where $\mathbf{A}$ is the vector potential associated with the incoming beam, $p$ is the polarization of the incoming beam whose value is either -1 for left circular polarization or 1 for right circular polarization, and where $m=l+p$ stands for z component the total angular momentum. The reduced rotation matrix $d_{mp}^j(\theta)$ can be found in \cite{Rose1955}, and the function $f_{q,l}(k_r)$ is related to the Fourier transform of the incident field, with $k_r$ being the transverse momentum, \textit{i.e.} $k_r=\sqrt{k_x^2+k_y^2}$. Note that the function $C_{jlp}$ is still defined as an integral, although it is no longer a double one. The reason why this happens is that the beams we have chosen as incident field have a specified value for $J_z$. Actually, eq. (\ref{general}) is valid for any beam whose $J_z$ is well defined. For the specific case of an incident paraxial LG$_{pl}$ beam, the integral in eq. (\ref{general}) can be solved analytically and the following expression is obtained:
\begin{equation}
\begin{split}
C_{jlp} = (-1)^l \left[ (j+p)!(j-p)!(j+l+p)!(j-l-p)! \right]^{1/2} \\
\sum_s \frac{(-1)^s}{(j-l-p-s)!(j+p-s)!(s+l)!s!} \frac{M_{l+2s}(f_{q,l})}{2^{l+2s}}
\end{split}
\label{Cjlp}
\end{equation}
where the number $M_a(f_{q,l})=\int_0^{\infty} k_r dk_r k_r^a f_{q,l}(k_r)$ is the momentum of order $a$ of the function $f_{q,l}$. The expression for $f_{q,l}(k_r)$ when the incident beam is a LG$_{q,l}$ is:
\begin{equation}
f_{q,l}(k_r)= \left( \frac{w_0^2q!}{2\pi (\vert l \vert + q )!} \right)^{1/2} \left(w_0 k_r /\sqrt{2} \right)^{\vert l \vert } L_q^{\vert l \vert}(w_0^2k_r^2/2)\exp \left( -w_0^2k_r^2/4 \right)
\label{F(LG)}
\end{equation}
with $w_0$ being the beam width of the LG mode in real space and $L_q^l(x)$ the associated Laguerre polynomials.
\begin{figure}[htbp]
\centering\includegraphics[width=7cm]{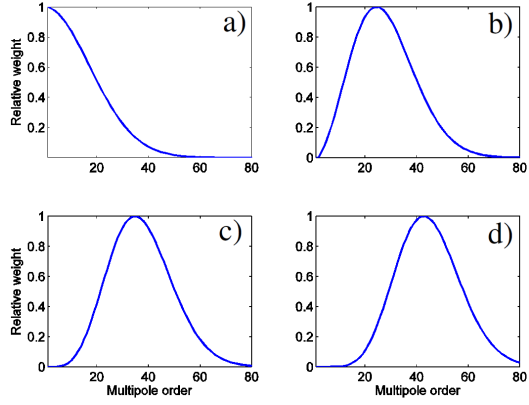}
\caption{$\vert C_{jlp} \vert^2$ for different LG$_{pl}$. a) LG$_{0,0}$ b) LG$_{0,2}$ c) LG$_{0,4}$ d) LG$_{0,6}$. The relation between the width ($w$) of the beam and the wavelength is $w/\lambda=4$.}
\end{figure}
Finally, for completeness sake, we give the expression of $M_a(f_{q,l})$ when the function $f_{q,l}$ is given by the expression (\ref{F(LG)}) and we also particularize the result for $q=0$: 
\begin{equation}
\begin{split}
M_a(f_{q,l})= \Gamma \left( \dfrac{a+\vert l \vert + 2}{2} \right) \sqrt{\frac{w_0^2 (\vert l \vert +q)!}{2\pi ( \vert l \vert ) !^2 q! }} \ 2^{a+\frac{\vert l \vert}{2}+1} \times \\ \times w_0^{-2-a} \ _2F_1\left( -q, \frac{a+\vert l \vert + 2}{2}, \vert l \vert +1; 2 \right) \\
\end{split}
\label{Ma}
\end{equation}
\begin{equation}
M_{l+2s}(f_{0,l})=  (s+ \vert l \vert )! w_0^{-1-\vert l \vert -2s}  \sqrt{\frac{1}{2\pi \vert l \vert!}} \  2^{\vert l\vert +2s+\frac{\vert l \vert}{2}+1}  
\label{Ma_LG}
\end{equation}
where $_2F_1(a,b,c:z)$ is the hypergeometric function. The problem, then, is solved. Using eqs. (\ref{general}), (\ref{Cjlp}) and (\ref{Ma_LG}) the decomposition of the incident beam is found and then the scattered and the interior fields are obtained by applying Maxwell boundary conditions. As a result, we obtain:
\begin{equation}
\begin{array}{ccl}
\textbf{A}^{sca}=&-\displaystyle\sum_{j=\vert l+p \vert}^{\infty} i^j (2j+1)^{1/2}  C_{jlp} \left[ b_j\textbf{A}_{j(l+p)}^{(m)}+ipa_j\textbf{A}_{j(l+p)}^{(e)} \right]  \\
\textbf{A}^{l}=&\displaystyle\sum_{j=\vert l+p \vert}^{\infty} i^j (2j+1)^{1/2} C_{jlp} \left[ c_j\textbf{A}_{j(l+p)}^{(m)}+ipd_j\textbf{A}_{j(l+p)}^{(e)} \right] \\
\end{array}
\end{equation}
where $C_{jlp}$ is given by eq. (\ref{Cjlp}) and its shape is depicted in Fig.1 for different cases. Also, $\left\lbrace a_n,b_n,c_n,d_n\right\rbrace$ are the classical Mie coefficients defined in \cite[Chap. 4]{Bohren1983}. Note that the two fields have the same formal expression as the incident field, except for the Mie coefficients and the radial dependence of the mulipolar fields, which is a Bessel function for the incident and interior fields and a Hankel function for the scattered one. These coefficients only depend on the radius of the particle ($R$), and the relative permeability ($\mu_r$) and permittivity ($\epsilon_r$) of the sphere with respect to the surrounding medium. Next, the analytical expressions for $a_j$ and $b_j$ are provided when $\mu_r=1$:
\begin{equation}
a_j=\frac{n_r\psi_j(n_r x)\psi_j'(x)-\psi_j(x)\psi_j'(n_r x)}{n_r\psi_j(n_r x)\xi_j'(x)-\xi_j(x)\psi_j'(n_r x)} \, \,  \quad \quad b_j=\frac{\psi_j(n_r x)\psi_j'(x)-n_r \psi_j(x)\psi_j'(n_r x)}{\psi_j(n_r x)\xi_j'(x)-n_r \xi_j(x)\psi_j'(n_r x)}
\label{mie_coeffs}
\end{equation}  
where $\psi_j$ and $\xi_j$ are the Riccati-Bessel functions of order $j$ and $x=2\pi r / \lambda$ is the so-called size parameter, and where $n_r=\sqrt{\mu_r \epsilon_r}$. 

\section{Cross sections}
Once the fields have been obtained, the scattering efficiency associated with them can be calculated. This efficiency is related to the power of a beam in the far field. Its definition, for the case of a plane wave excitation, is given by the flux of the Poynting vector across a spherical surface divided by the product of the incident irradiance and the particle's cross-sectional area projected onto a plane perpendicular to the incident beam \cite[Chap. 3]{Bohren1983}:  
\begin{equation}
Q^{sca}_{Mie}= \frac{1}{I_i \cdot \pi R^2}\int \textbf{S}^{sca}\cdot d\Omega 
\label{Poynting}
\end{equation}
with $\textbf{S}^{sca}= \text{Re} \left\lbrace (1/2) \textbf{E}^{sca} \times {\textbf{H}^{sca}}^{*} \right\rbrace$ the Poynting vector, $d\Omega$ an element of surface perpendicular to the surface, and $I_i$ the incident irradiance. This expression has a major drawback when it is applied to incident fields that vary point to point. Hence, we have used another normalization factor to compute the scattering efficiency. We normalized the expressions over the integral of $\vert \mathbf{E}^i \vert^2$ across a transverse surface:
\begin{equation}
Q^{sca}= \frac{1}{\int \vert \mathbf{E}^i \vert^2 d\Omega}\int \textbf{S}^{sca}\cdot d\Omega 
\label{Poynting}
\end{equation}
To perform this calculation, we must derive the electric and magnetic fields from the vector potential in a Coulomb gauge, that is:
\begin{eqnarray}
\mathbf{E}&=&ik\mathbf{A}\\
\mathbf{H}&=&\nabla \times \mathbf{A}
\end{eqnarray}
Then, the integral (\ref{Poynting}) is performed paying special attention to the orthonormality relations of the multipolar fields, which are:
\begin{eqnarray}
\int \left(\textbf{A}_{jm}^{(m)} \times {\textbf{A}_{jm}^{(m)}}^* \right) \cdot \textbf{r} \, \,  d\Omega = 0 \\
\int \left(\textbf{A}_{jm}^{(e)} \times {\textbf{A}_{jm}^{(e)}}^* \right) \cdot \textbf{r} \, \,  d\Omega = 0 \\
\int d\Omega \, \,\textbf{A}_{jm}^{(x')} {\textbf{A}_{jm}^{( x' )}}^*=F\left( r \right) \delta_{jj'} \delta_{mm'} \delta_{xx'}
\end{eqnarray}
where $F(r)$ is 
\begin{equation}
F(r)= \left\lbrace \begin{array}{cccl} \vert \xi_j \vert^2 & \text{for} & x=m \\ \dfrac{j \vert \xi_{j+1} \vert^2 + (j+1) \vert \xi_{j-1} \vert^2}{2j+1} & \text{for} & x=e  \end{array} \right. 
\end{equation}
and $\xi_j$ is a Bessel function for the incident and interior field and a Hankel function of the first kind for the scattered field.\\
Finally, the scattering efficiency for the incident beam in eq. (\ref{general}) is:
\begin{equation}
Q^{sca}=\sum_{j={ \vert l+p \vert}}^{\infty} \frac{2 \pi \left( 2j+1 \right)}{k^2} \vert C_{jlp} \vert ^2 \left( \vert a_j \vert ^2 + \vert b_j \vert ^2 \right) 
\label{Csca}
\end{equation}
with $C_{jlp}$ given by eq. (\ref{Cjlp}) and where the following relation holds $\sum_{j}(2j+1)\vert C_{jlp} \vert ^2=2$ if the incident field is normalized to 1 \cite{Gabi2008}. In contradiction, the result from the Mie theory [7] is:
\begin{equation}
Q_{Mie}^{sca}=\sum_{j={1}}^{\infty} \frac{ \left( 2j+1 \right)}{x^2} \left( \vert a_j \vert ^2 + \vert b_j \vert ^2 \right) 
\label{CscaMie}
\end{equation}
Expressions (\ref{Csca}) and (\ref{CscaMie}) give rise to very different scattered cross sections. We show these differences in Fig. 2, 3 and 4. First, expression (\ref{CscaMie}) is used to compute the Mie scattering cross section. The result of this computation is depicted in Fig. 2. It can be seen that the scattering cross section behaviour is dominated by the low order modes and that the ripples are caused by the high order electromagnetic modes. 
\begin{figure}[htbp]
\centering\includegraphics[width=7cm]{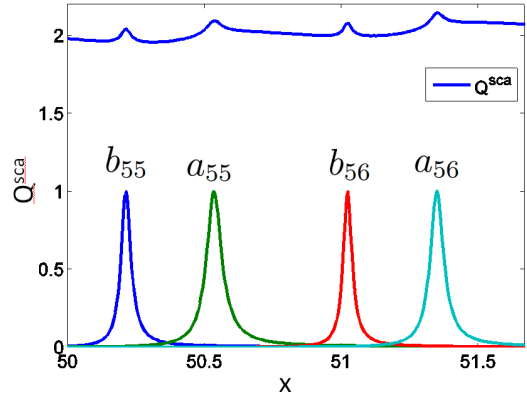}
\caption{$Q^{sca}$ and some of the Mie coefficients for the conventional Mie Theory, when $m=1.33+i 10^{-8}$ and $x \in \left[50,52  \right]$  (Ref. \cite{Bohren1983}, Fig. 11.7).}
\end{figure}
Next, expression (\ref{Csca}) is used to plot Fig. 3 and Fig. 4. In Fig. 3 the scattering cross section is plotted for four different LG$_{q,l}$ modes. It can be observed that the angular momentum of the incident beam plays a crucial role. Indeed, the ripple structure is increasingly enhanced as the angular momentum of the LG beam in consideration is increased. This effect is specially clear in Fig. 3d where some of the ripples present in Fig. 3a are largely enhanced. 
\begin{figure}[htbp]
\centering\includegraphics[width=9cm]{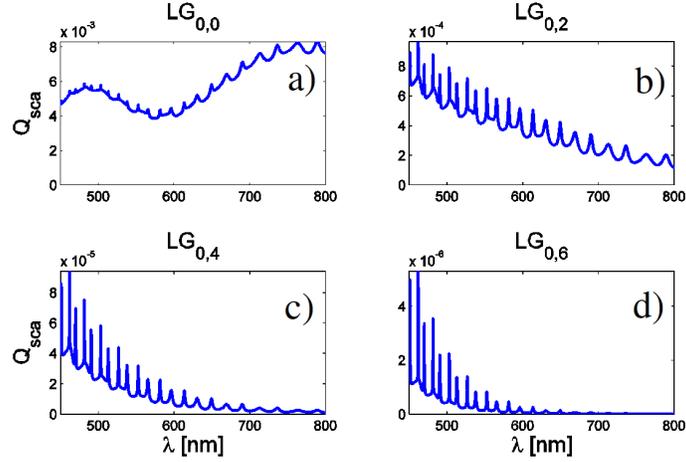}
\caption{$Q^{sca}$ as a function of the wavelength for different incident LG beams. a) LG$_{0,0}$ b) LG$_{0,2}$ c) LG$_{0,4}$ d) LG$_{0,6}$. The other parameters are kept equal to $n_r=1.5$, $p=1$, $r=1.3 \ \mu m$. The decomposition of each of the LG beams used to computed these cross sections is shown in Fig. 1.}
\end{figure}
Finally, Fig. 4 shows the enhancement of a single ripple. 
\begin{figure}[htbp]
\centering\includegraphics[width=9cm]{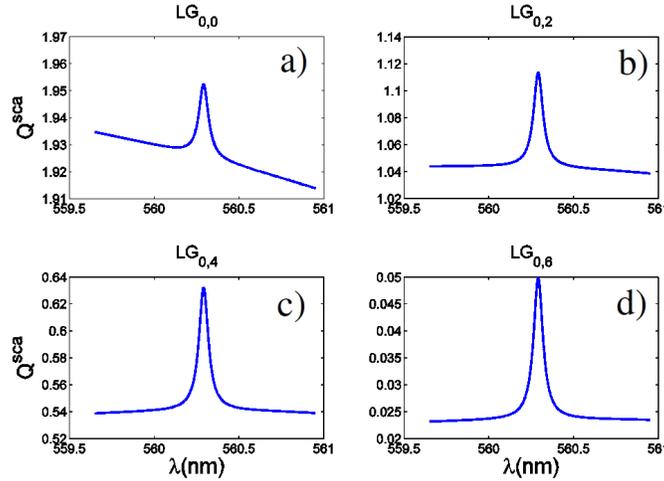}
\caption{$Q^{sca}$ for a very short range of wavelegths. a) LG$_{0,0}$ b) LG$_{0,2}$ c) LG$_{0,4}$ d) LG$_{0,6}$. Equally to Fig. 3, the other variables are kept equal to $n_r=1.5$, $p=1$, $r=1.3 \ \mu m$.}
\end{figure}
Although presented in a different way, the same effect presented in Fig. 3 is observed again. However, in this case, the enhancement of the ripple structure can be determined quantitatively. We have compared the maximum value of the cross section which peaks at 560.3 nm and the minimum value. This minimum value has been computed as the average of two values at 0.7nm apart in wavelength. Then, we have computed the ratio $\gamma$, which is the relative quotient between the subtraction of the maximum and the minimum, and the minimum.
\begin{equation}
\gamma= \dfrac{Q_{max}-Q_{min}}{Q_{min}} 
\label{gamma}
\end{equation}
where $Q_{max}=Q^{sca}_{560.3}$ and $Q_{min}=(Q^{sca}_{561.0}+Q^{sca}_{559.6})/2$. The results are presented in Table 1. It can be seen that the background is highly reduced. When a  LG$_{0,0}$ is used to excite the sphere, the peak only represents 1$\%$ of the background. Nonetheless, when a LG$_{0,6}$ is used, the peak contribution is larger than the background's one. That is, the background has been reduced more than a half. 
\begin{table}  
\begin{center}  
\begin{tabular}{| l | l | l | l | l |}
\hline Fig. 4 & LG$_{0,0}$ & LG$_{0,2}$ & LG$_{0,4}$ & LG$_{0,6}$ \\
\hline $\gamma (\%)$ & 1.455 & 6.916 & 17.30 & 113.6 \\
\hline
\end{tabular}
\caption{Enhancement of the ripple structure using different LG beams. }
\label{table}
\end{center}
\end{table}

\section{Conclusions}
We have shown that the LG beams can be used to enhance the ripple sctructure in dielectric spheres. The conservation of the AM of light implies that the first multipolar modes do not contribute to the scattering of the particle and therefore the ripple structure can be turned into resonances. This is a great improvement towards the control of scattered fields. This new effect could be applied in a large variety of fields such as spectroscopy, cytometry or dark-field microscopy to mention a few.

\section*{Acknowledgements}
We thank Xavier Vidal for his suggestions during the writing of this article. This work was funded by the Australian Research Council Discovery Project DP110103697. G.M.-T. is the recipient of an Australian Research Council Future Fellowship (project number FT110100924)

%\bibliographystyle{elsarticle-num-names}
%\bibliography{xavizpcite}

\end{document}